\documentclass[a4paper]{article}
\usepackage{Odyssey2018}
\usepackage{epsfig,amssymb,amsmath}
\usepackage{amsmath,graphicx}
\usepackage{multirow}
\usepackage{subfigure}
\usepackage{algorithm}
\usepackage{algorithmicx}
\usepackage{algpseudocode}
\usepackage{bm}
\usepackage{enumerate}

\ninept
\title{\center   Exploring the Encoding Layer and Loss Function in 
	End-to-End \\ Speaker and Language Recognition System }

\name{Weicheng Cai$^{2}$, Jinkun Chen$^2$ and Ming Li$^{1}$}

\address{$^1$Data Science Research Center, Duke Kunshan University, Kunshan, China\\
	$^2$School of Electronics and Information Technology, Sun Yat-sen University, Guangzhou, China\\
	{\small \tt ming.li369@dukekunshan.edu.cn}}

\begin{document}
	\maketitle

	\begin{abstract}
	In this paper, we explore the encoding/pooling layer and loss function in the end-to-end speaker and language recognition system.  First,  a unified and interpretable end-to-end system for both speaker and language recognition is developed. It accepts variable-length input and produces an utterance level result.  In the end-to-end system, the encoding layer plays a role in aggregating the variable-length input sequence into an utterance level representation. Besides the basic temporal average pooling, we introduce a self-attentive pooling layer and a learnable dictionary encoding layer to get the utterance level representation. In terms of loss function for open-set speaker verification, to get more discriminative speaker embedding, center loss and angular softmax loss is introduced in the end-to-end system. Experimental results on  Voxceleb and NIST LRE 07 datasets show that the performance of end-to-end learning system could be significantly improved by the proposed encoding layer and loss function.
	\end{abstract}

	\section{Introduction}

	Language recognition (LR) , text-independent speaker recognition (SR) and many other paralinguistic speech attribute recognition tasks can be defined as an utterance level ``sequence-to-one" learning issue, compared with automatic speech recognition, which is a ``sequence-to-sequence" tagging task. They are problems in that we are trying to retrieve information about an entire  utterance rather than specific word content \cite{campbell2006support}. Moreover, there is  no constraint on the lexicon words thus the training utterances  and testing segments may have completely different contents \cite{Kinnunen2010An}. The goal, therefore, may boil down to find a robust and time-invariant utterance level vector representation describing the distribution of the given input  data sequence with variable-length. 
	
\begin{figure*}[tb]
	\centering
	\includegraphics[width=0.96\textwidth]{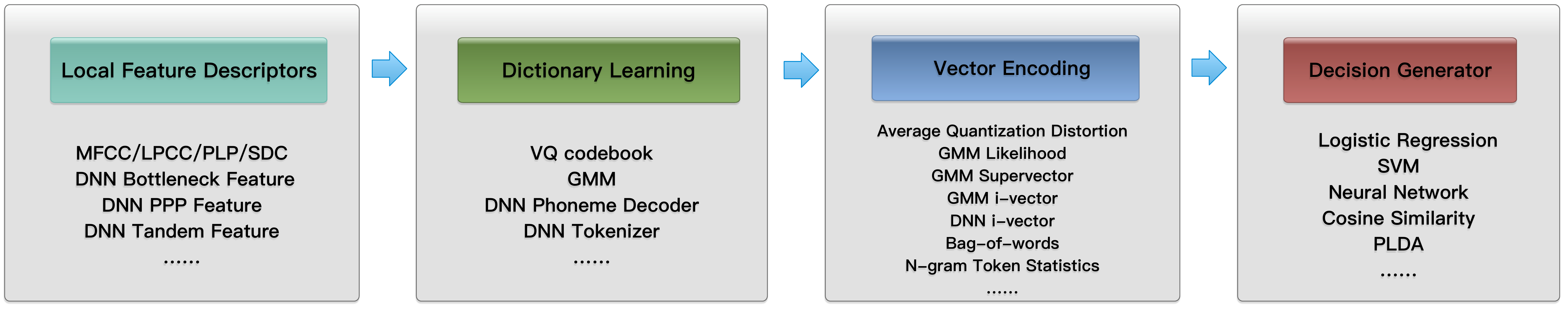}
	\caption{Four main steps in the conventional processing pipeline}\label{fig:offtheshelf}
\end{figure*}

In recent decades, the classical GMM i-vector approach and its variants  have dominated multiple kinds of paralinguistic speech attribute recognition fields for its superior performance, simplicity and efficiency \cite{dehak2010front,Dehak2011Language}. As shown in Fig. \ref{fig:offtheshelf},  the conventional processing pipeline contains four main steps as follows:

	\begin{itemize}
		\setlength{\itemsep}{0pt}
		\setlength{\parsep}{0pt}
		\setlength{\parskip}{0pt}
		\item Local feature descriptors,  which manifest as a variable-length feature sequence, include hand-crafted acoustic level features, such as log mel-filterbank energies (Fbank), mel-frequency cepstral coefficients (MFCC), perceptual linear prediction (PLP), shifted delta coefficients (SDC) features \cite{Kinnunen2010An,6451097}, and automatically learned phoneme discriminant features from deep neural networks (DNN), such as bottleneck features \cite{matejka2014neural, Song2015Deep,richardson2015unified}, phoneme posterior probability (PPP) features \cite{Li2016Generalized}, and  tandem features \cite{li2014interspeech, Richardson2015Deep}. 
		
		\item Dictionary, which contains several temporal orderless center components (or units, words, clusters, etc.), includes  vector quantization (VQ) codebooks learned by K-means \cite{Soong1985Report}, a universal background model (UBM) learned by Gaussian Mixture Model (GMM) GMM \cite{Reynolds1995Robust,Reynolds2000Speaker} or a supervised phonetically-aware acoustic model learned by DNN \cite{li2014interspeech,yun_icassp14}.
		\item Vector encoding. This procedure  aggregates the  variable-length feature sequence  into an utterance level  vector representation,  based on the statistics learned on the dictionaries mentioned above. Typical examples are the GMM Supervector/i-vector \cite{campbell2006support,dehak2010front} or the recently proposed DNN i-vector \cite{yun_icassp14, Snyder2016Time}.
		\item Decision generator, includes logistic regression (LogReg),  support vector machine (SVM), and neural network for closed-set identification, cosine similarity or probabilistic linear discriminant analysis (PLDA) \cite{Prince2007Probabilistic,Kenny2010Bayesian} for open-set verification.
	\end{itemize}

\begin{figure}[tb]
	\centering
	\subfigure[Closed-set identification]{
		\label{fig:subfig:a} 
		\includegraphics[width=0.32\textwidth]{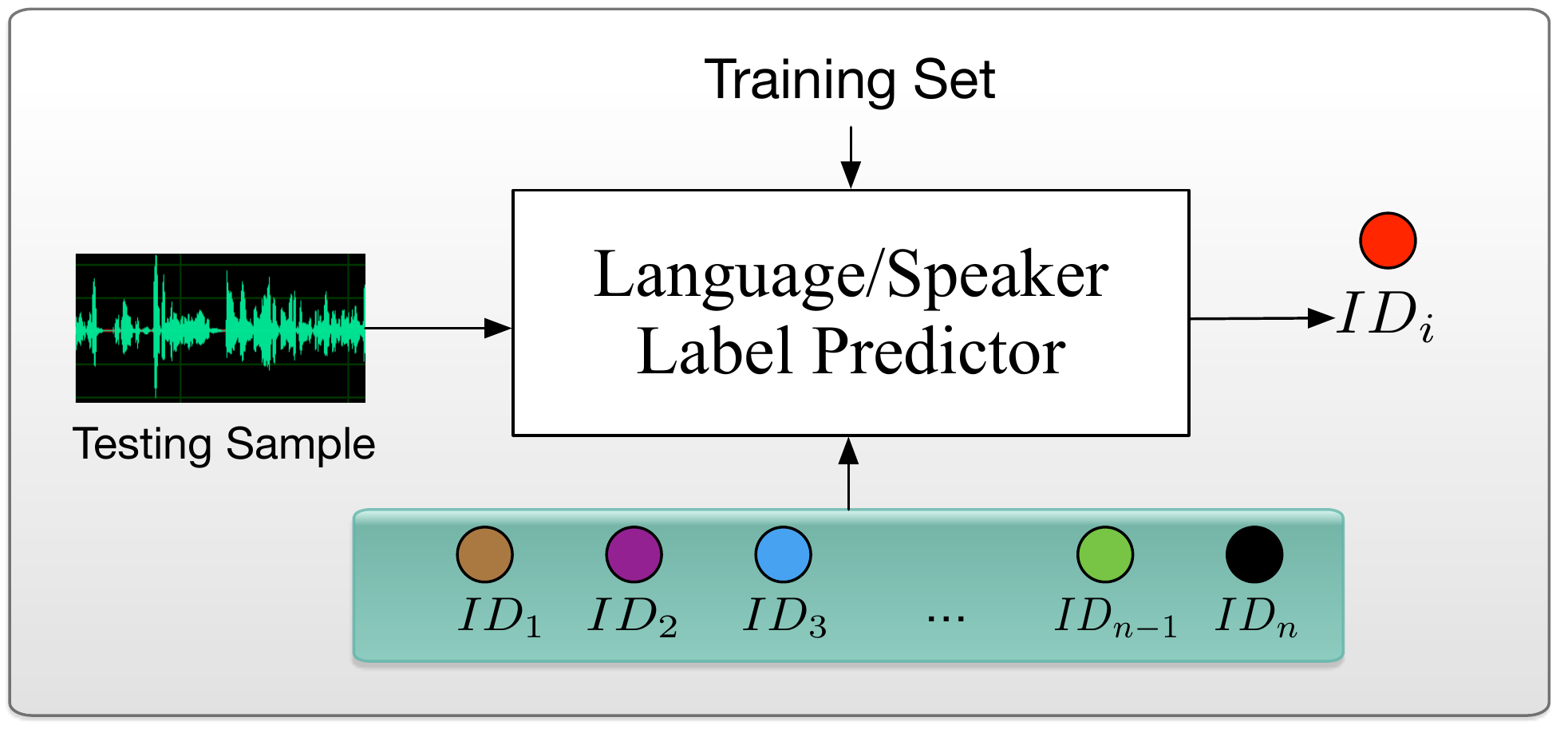}}
	\hspace{0.01in}
	\subfigure[Open-set verification]{
		\label{fig:subfig:b} 
		\includegraphics[width=0.32\textwidth]{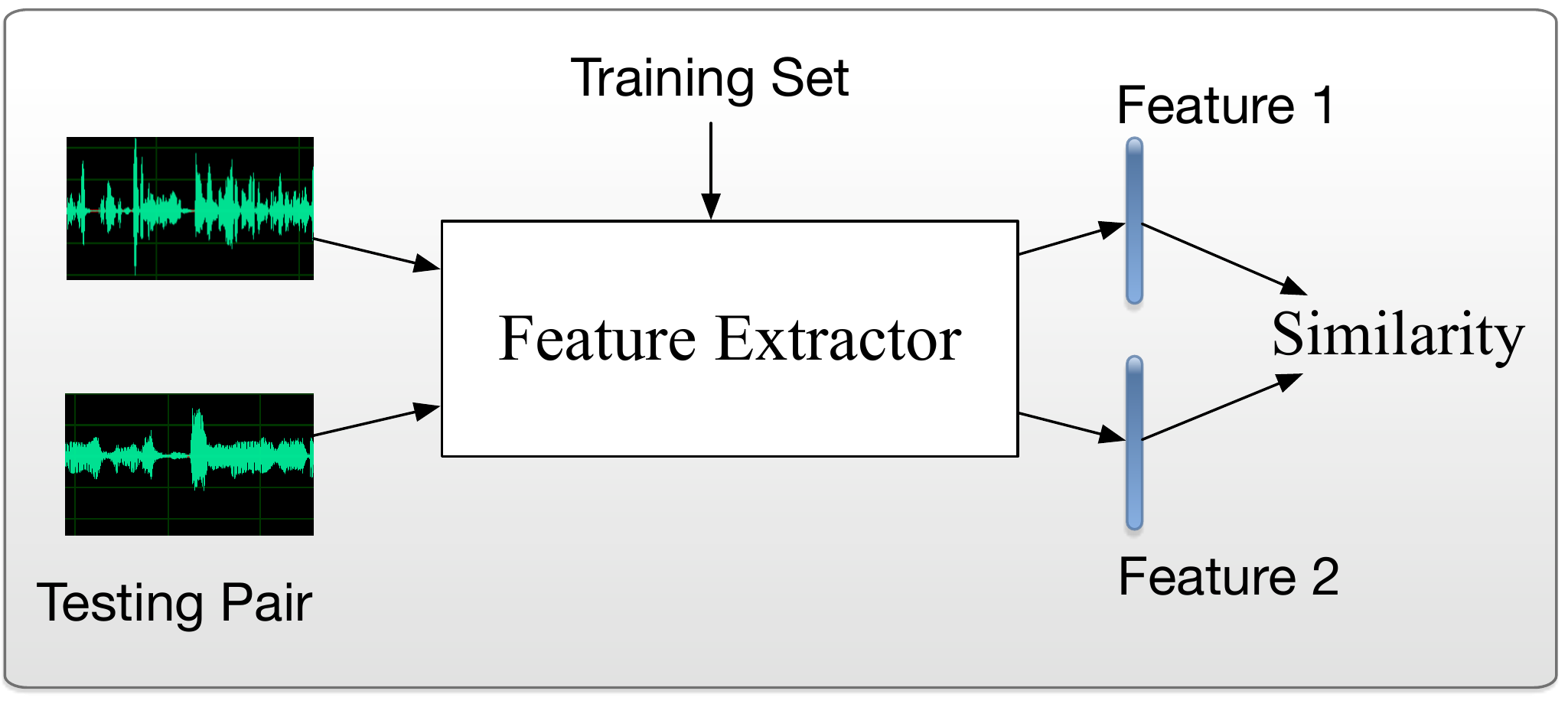}}
	\caption{Comparison of closed-set identification and open-set verification problem. The closed-set identification is equivalent to classification task, while the open-set verification can be considered as a metric learning task}\label{fig:definition}
\end{figure}
 
	The GMM i-vector based approaches comprise a series of hand-crafted or ad-hoc algorithmic components, and they show strong generalization ability and robustness when data and computational resource are limited.  In recent years, with the merit of large labeled datasets, enormous computation capability, and effective network architectures, emerging progress towards end-to-end learning opens up a new area for exploration \cite{lopez2014automatic,gonzalez2014automatic,Snyder2017Deep,1705.02304,jin2018lid}.  In our previous works \cite{caie2e_iccasp18, cailde_iccasp18}, we proposed a learnable dictionary encoding (LDE) layer, which connects the conventional GMM Supervector procedure  and state-of-the-art end-to-end neural network together. In the end-to-end learning scheme, a general encoding layer is employed on top of the front-end convolutional neural network (CNN), so that it can encode the variable-length input sequence  into an utterance level representation  automatically. We have shown its success for closed-set LR task. However, when we move forward to SR task, the situation becomes much more complicated.  

	Typically, SR can be categorized as speaker identification
	and speaker verification. The former classifies a speaker to a specific identity, while the latter determines whether a pair of utterances belongs to the same person. In terms of the testing protocol, SR can be evaluated under closed-set or open-set settings, as illustrated
	in Fig. \ref{fig:definition}.  For closed-set protocol, all testing identities are
	enrolled in the training set. It is natural to classify a testing utterance to a given identity. Therefore, closed-set
	language or speaker identification can be well addressed as a classification problem. For the open-set
	protocol, speaker identities in testing set are usually disjoint from the ones in 
	training set, which makes the speaker verification  more challenging yet closer to
	practice. Since it is impossible to classify testing utterances to known
	identities in training set, we need to map speakers to a discriminative
	feature space. In this scenario, open-set speaker verification is essentially a metric learning problem, where
	the key  is to learn discriminative large-margin features.
	
	Considering the aforementioned challenges,  we generalize the learning scheme for closed-set LR in \cite{caie2e_iccasp18}, and build a unified end-to-end system for both LR and SR. The whole pipeline contains five key modules:  input data sequence, frame-level  feature extractor, encoding layer, loss function, and similarity metric. In this paper, We focus on  investigating  how to enhance the system performance by exploring different kinds of  encoding layers and loss functions. 
	
	\section{End-to-End System Overview}
	
\begin{figure*}[htbp]
		\centering
		\includegraphics[width=0.95\textwidth]{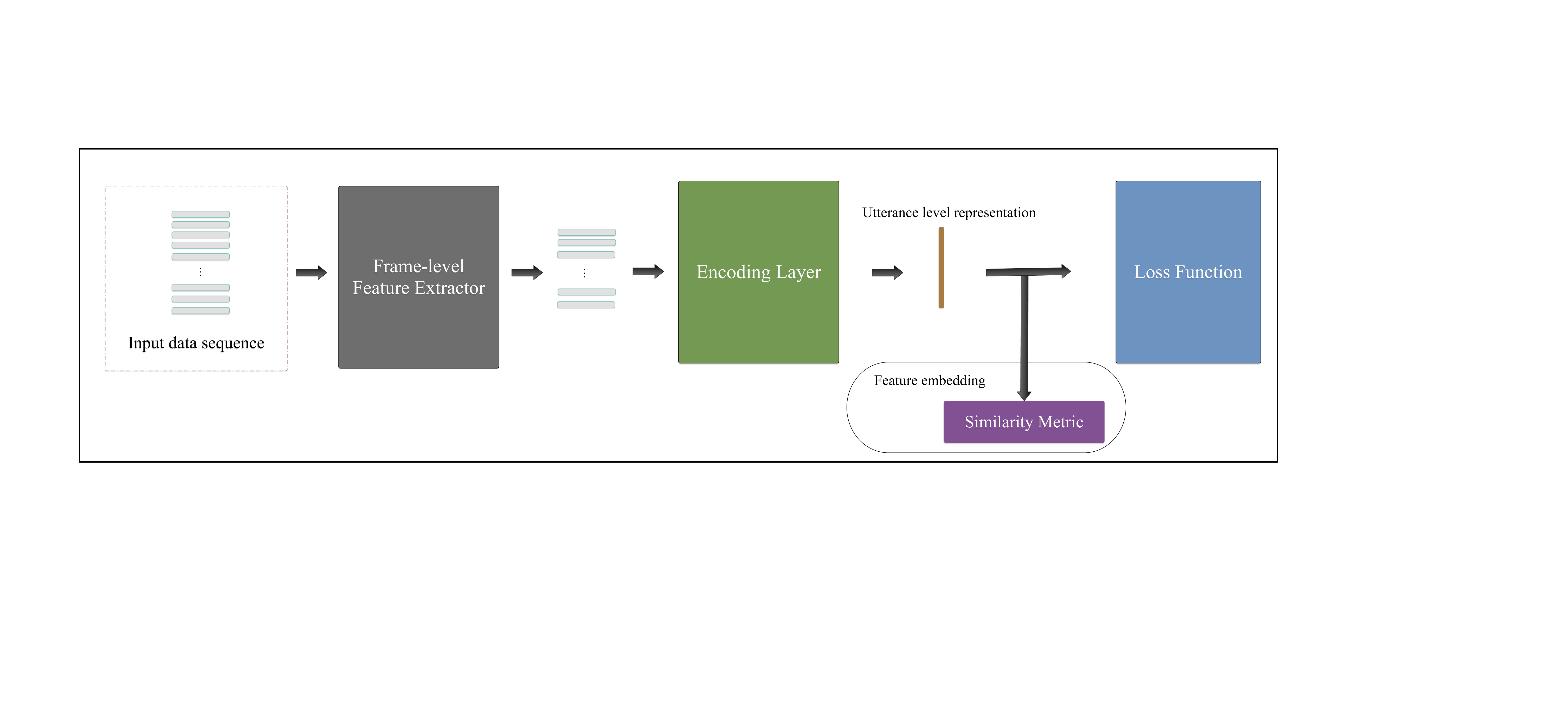}
		\caption{End-to-end framework for both LR and SR. It accepts input data sequence with variable length, and  produces an utterance level result. The whole pipeline contains five key modules:  input data sequence, frame-level  feature extractor, encoding layer, loss function, and similarity metric. The additional similarity metric module is specifically designated for the open-set  verification task.  }\label{fig:format}
	\end{figure*}

	The speech signal is naturally with variable length, and we usually don't know exactly how long the testing speech segment will be. Therefore, a flexible processing method should have the ability to accept speech segments with arbitrary duration. Motivated by \cite{Snyder2017Deep,1705.02304,caie2e_iccasp18}, the whole end-to-end framework in this paper is shown in Fig. \ref{fig:format}. It accepts variable-length input and produces an utterance level result. The additional similarity metric module is specifically designated for the open-set  verification task. 
	
Given input data feature sequence such as log mel-filterbank energies (Fbank),  we employ a deep convolutional neural network (CNN)  as our frame-level feature extractor. It can learn high-level abstract local patterns from the raw input automatically.  The frame-level representation after the front-end convolutional layers is still in a temporal order. The remaining issue is to aggregate them together over the entire sequence. In this way, the encoding layer plays a role in extracting a fixed-dimensional utterance level representation from a variable-length input sequence.  The utterance level representation is further processed through a  fully-connected (FC) layer and finally connected with an output layer. Each unit in the output layer is represented as a target speaker/language label. All the components in the pipeline are jointly learned in an end-to-end manner with a unified loss function.

	


	\section{Encoding layer}

	\subsection{Temporal average pooling layer}
	Recently, in both \cite{Snyder2017Deep, 1705.02304}, similar temporal average pooling (TAP) layer is adopted in their neural network architectures. As shown in Fig. \ref{fig:encoding},  the TAP layer is inherently designated in the end-to-end network, and it equally pools the front-end  learned features over time.

\subsection{Self-attentive pooling  layer}
	The TAP layer equally pools the  CNN extracted features over time. However, not all frame of features contribute equally to the utterance level representation, We introduce a self-attentive pooling (SAP) layer to pay attention to 
such frames that are important to the classification and aggregate those informative frames to form a utterance level representation.  

In \cite{geng2016end}, attention-based recurrent neural network (RNN) is introduced to get utterance level representation for closed-set LR task . However, the work in \cite{geng2016end} relies on a non-trivial pre-training procedure to get the language category embedding, and the authors only report results on 3s short duration task. Different from \cite{geng2016end} , the attention mechanism in our  network architecture is self-contained, with no need for extra guiding source information. 

We implement the SAP layer similar to \cite{yang2016hierarchical,bhattacharya2017deep,chowdhury2017attention}. That is, we first feed the utterance level feature maps $\{\bm{x_{1}},\bm{x_{2}},\cdots,\bm{x_{L}}\}$  into 
a multi-layer  perceptron (MLP)  to get $\{\bm{h_{1}},\bm{h_{2}},\cdots,\bm{h_{L}}\}$ as a hidden representation. In this paer, we simply adopt  a  one-layer perceptron, 
\begin{equation}
\bm{h}_{t} = \tanh(\bm{Wx}_{t} + \bm{b})
\end{equation}
Then we measure the importance of each frame as the similarity of $\bm{h_{t}}$ with a learnable 
context vector $\bm{\mu}$ and get a normalized importance
weight $w_t$ through a softmax function.
\begin{equation}
w_{t} = \frac{\exp(\bm{h}_{t}^T\bm{u})}{\sum_{t=1}^{T}\exp({\bm{h}_{t}^T}\bm{u})}
\end{equation}

The context vector $\bm{\mu}$
can be seen as a high level representation of a fixed
query ``what is the informative frame” over the whole frames \cite{yang2016hierarchical}. It is randomly initialized and jointly learned
during the training process.

After that,  the utterance level representation $\bm{e}$  can be generated as a weighted sum of the frame level CNN feature maps
based on the learned weights. 
\begin{equation}
\bm{e} = \sum_{t=1}^{T}w_t \bm{x}_{t}
\end{equation}

\subsection{Learnable dictionary encoding layer}

 In conventional speaker verification system, we always rely on a dictionary learning procedure like K-means/GMM/DNN, to accumulate statistics. Inspired by this, we introduce a novel LDE Layer to accumulate statistics on more detailed units. It combines the dictionary learning and vector encoding steps into a single layer for end-to-end learning. 
	
\begin{figure*}[tb]
		\centering
		\includegraphics[width=0.92\textwidth]{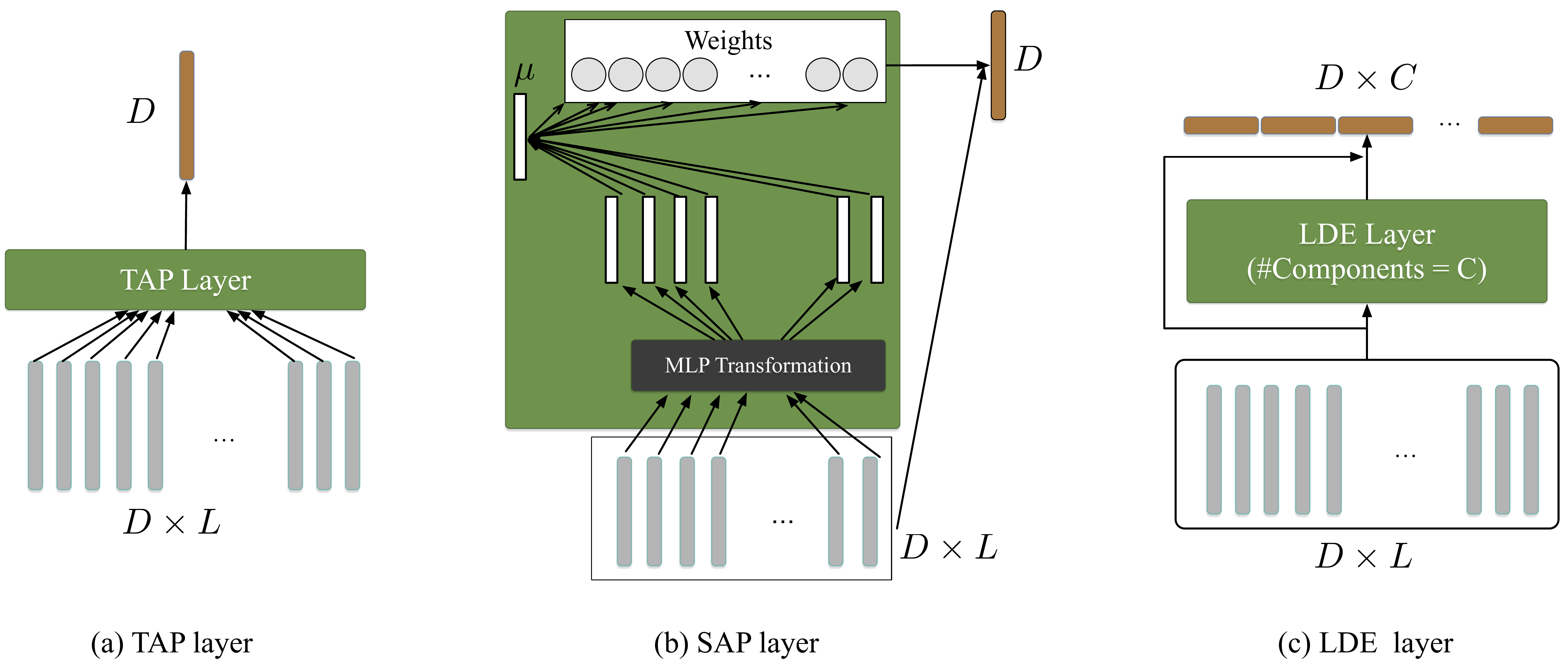}
		\caption{Comparison of different encoding procedures}\label{fig:encoding}
	\end{figure*}

As demonstrated in Fig. \ref{fig:encoding},  given an input temporal ordered feature sequence with the size of $D \times L$ (where $D$ denotes the feature coefficients dimension, and $L$ denotes the temporal duration length), LDE layer aggregates them over time. More specifically, it transforms them into an utterance level temporal orderless $D \times C$ vector representation, which is independent of length $L$.  The LDE Layer imitates the mechanism  of GMM Supervector, but learned directly from the loss function.
	
\begin{figure}[tb]
			\centering
			\includegraphics[width=0.35\textwidth]{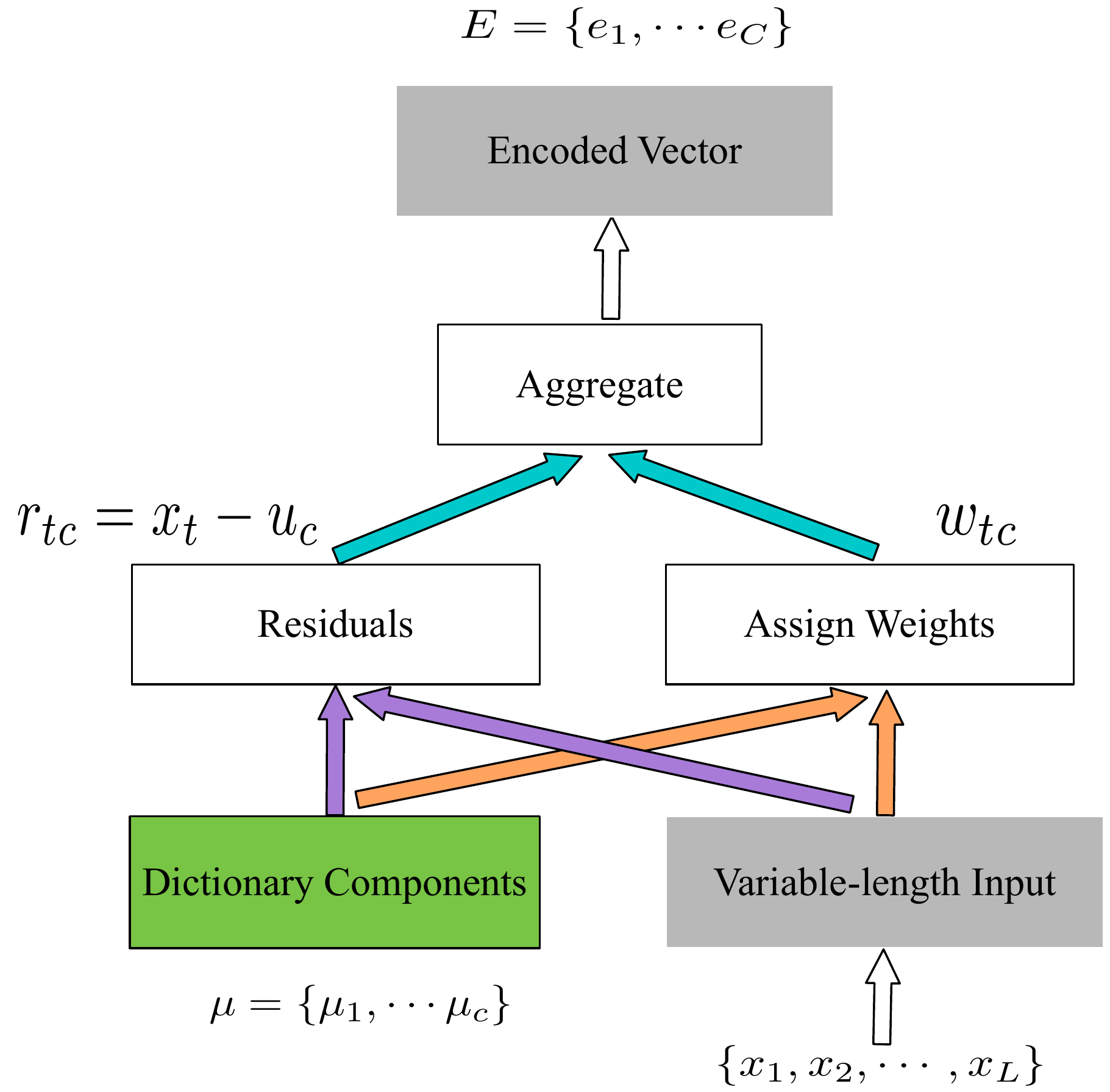}
			\caption{The forward diagram within the  LDE layer }\label{fig:forward_diagram}
		\end{figure}

The LDE layer is a directed acyclic graph and all the components are differentiable $w.r.t$ the input $\bm{X}$ and the learnable  parameters. Therefore, the LDE layer can be trained in an end-to-end manner by standard stochastic gradient descent with backward propagation. Fig. \ref{fig:forward_diagram} illustrates the forward diagram of LDE layer. Here, we introduce two groups of learnable  parameters. One is the  dictionary component center, noted as $\bm{\mu} = \{\bm{\mu_1},  \bm{\mu_2} \cdots \bm{\mu_c} \}$. The other one is assigned weights,  noted as $\bm{w}$.
	
	 Consider  assigning weights from the features to the dictionary components. Similar as soft-weight assignment in GMM,  the features are independently assgined to each dictionary component and  the non-negative assigning weight is given by a softmax function,

	\begin{equation}
	w_{tc} = \frac{\exp(-s_c\Vert \bm{{r}_{tc}}\Vert ^2)}{\sum_{m=1}^{C}\exp(-s_m\Vert \bm{{r}_{tm}}\Vert ^2)}
	\end{equation}
where the smoothing factor $s_c$ for each dictionary center $\bm{u_c}$ is learnable.

	Given a set of $L$ frames feature  sequence $\{\bm{x_{1}},\bm{x_{2}},\cdots,\bm{x_{L}}\}$ and a learned dictionary center $\bm{\mu} = \{\bm{\mu_1},  \bm{\mu_2} \cdots \bm{\mu_c} \}$, each frame of feature $\bm{x_t}$ can be assigned with a weight $w_{tc}$ to each component $\bm{{\mu_c}}$ and the corresponding residual vector is denoted by $\bm{r_{tc}} = \bm{x_{t}}  -  \bm{u_{c}}$, where $t = 1,2\cdots L$ and $c = 1, 2\cdots C$. Given the assignments and the residual vector, similar to conventional GMM Supervector, the residual encoding model applies an aggregation operation for every dictionary component center $\bm{\mu_c}$:
	\begin{equation}
	\bm{e_{c}}=\sum_{t=1}^{L}e_{tc}=\frac{\sum_{t=1}^{L}(w_{tc} \bm{\cdot} \bm{r_{tc}})}{\sum_{t=1}^{L}w_{tc}}
	\label{eq_2}
	\end{equation}

In order to facilitate the derivation， we simplified it as 
	\begin{equation}
	\bm{e_{c}}=\frac{\sum_{t=1}^{L}(w_{tc} \bm{\cdot} \bm{r_{tc}})}{L}
	\label{eq_2}
	\end{equation}

	The LDE layer concatenates the aggregated
	residual vectors with assigned weights. The resulted encoder outputs a fixed dimensional representation $\bm{E} = \{\bm{e_1} , \bm{e_2} \cdots \bm{e_C} \}$.

	\section{Loss function}
	\subsection{Loss function for closed-set identification}
	In conventional LR or SR problem, the processing stream is explicitly separated into front-end and back-end. The i-vector extracting front-end is comprised of multiple unsupervised generative models. They are optimized through Expectation Maximum (EM) algorithm under a  negative complete-data log-likelihood loss. Since they are all generative models, we refer their loss functions as a kind of generative negative log-likelihood (GNLL) loss for simplicity. Once front-end model is trained and i-vector is extracted, a back-end LogReg  or SVM is commonly adopted to do the back-end classification. Their loss function is softmax or hinge loss.

	As illustrated in Fig. \ref{fig:loss}, for an end-to-end closed-set identification system, the front-end feature extractor and back-end classifier could be jointly learned. In this way, the whole identification system could be optimized within a unified softmax loss:
	
	\begin{equation}
	\ell_{s}=-\frac{1}{M}\sum_{i=1}^M \log\frac{e^{\bm{W}_{y_i}^Tf(\bm{x}_i)+\bm{b}_{y_i}}}{\sum_{j=1}^{C}e^{\bm{W}_j^Tf(\bm{x}_j)+\bm{b}_j}} 
	\end{equation}
	where $M$ is the training batch size, $\bm{x_i}$ is the $i^{th}$ input data sequence in the batch, $f(\bm{x_i})$ is the corresponding output of the penultimate layer of the end-to-end neural network, $y_i$ is the corresponding target label, and $\bm{W}$ and $\bm{b}$ are the weights and bias for the last layer of the network which acts as a classifier.

	\subsection{ Loss function for open-set verification}
	Once front-end model is trained and i-vector is extracted,  PLDA is commonly adopted in the state-of-the-art  open-set speaker verification system. PLDA is a Bayesian generative model. Thus its loss function is still GNLL.

\begin{figure}[tb]
		\centering
		\subfigure[Closed-set identification]{
			\label{fig:subfig:a} 
			\includegraphics[width=0.98\columnwidth]{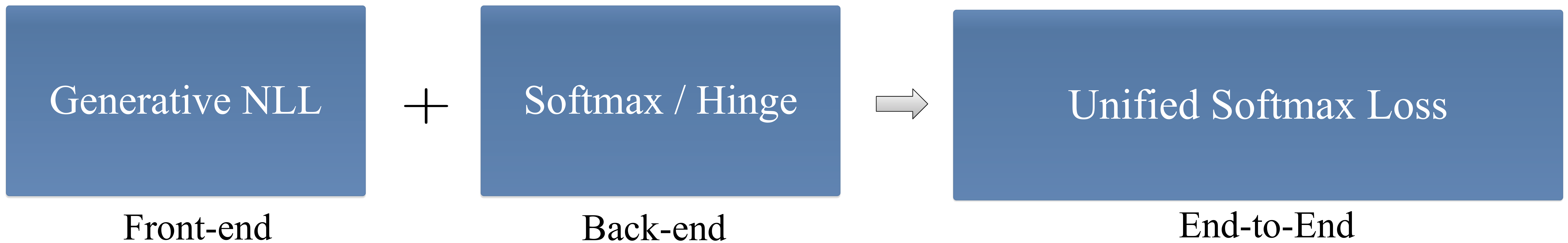}}
		\hspace{0.01in}
		\subfigure[Open-set verification]{
			\label{fig:subfig:b} 
			\includegraphics[width=\columnwidth]{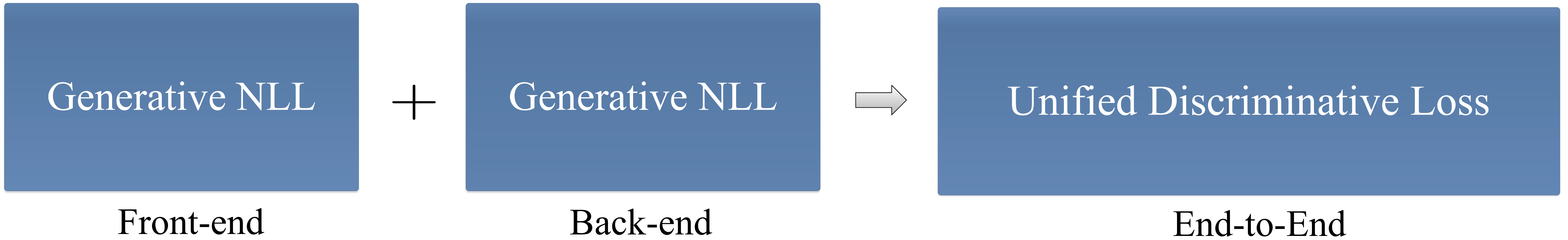}}
		\caption{Conventional explicitly separated front-end and back-end loss are proceeded  into a unified end-to-end loss }\label{fig:loss}
	\end{figure}

	We believe that PLDA is not necessary, and a completely end-to-end system should have ability to learn this kind of open-set problem with a unified loss function. However, for open-set speaker verification task, the learned feature embedding  need to be not only separable but also discriminative. Since it is
	impractical to pre-collect all the possible testing identities for
	training, the label prediction goal and corresponding basic softmax loss is not always applicable. Therefore, as illustrated in Fig. \ref{fig:loss}, a unified discriminative loss function is needed to have better generalization than closed-set identification:

 In \cite{Snyder2017Deep, 1705.02304}, similar pairwise loss such as contrastive loss \cite{Hadsell2006Dimensionality,Chen2014Deep} or triplet loss \cite{Schroff2015FaceNet} is adopted for open-set speaker verification. They all explicitly treat the open-set speaker verification problem as metric learning problem. However, a neural network trained with pairwise loss requires carefully designed pair/triplet mining procedure. This procedure is non-trivial,  both time-consuming and performance-sensitive \cite{liu2017sphereface}. In this paper, we focus on the general classification network. This means the units in the output layer are equal to the speaker numbers in the training set. Here we introduce two discriminative loss which is first proposed in computer vision community.

\subsubsection{Center loss}
 The basic softmax loss encourages the separability of features only. In \cite{wen2016discriminative}, the authors propose a center loss simultaneously learning a center for deep
	features of each class and penalizing the distances between the
	deep features and their corresponding class centers.
	The learning goal is to minimize the within-class variations
	while keeping the features of different classes separable. The joint supervision of softmax loss and center
	loss  is adopted for discriminative feature learning:
	\begin{scriptsize}
	\begin{equation}
	\begin{split}
	&\ell= \ell + \lambda \ell_C  \\
	&= -\frac{1}{M}\sum_{i=1}^M log \frac{e^{\bm{W}_{y_i}^T f(\bm{x}_i) + \bm{b}_{y_i}}}{\sum_{j=1}^C e^{\bm{W}_j^Tf(\bm{x}_i) + \bm{b}_j}} + \frac{\lambda}{2}\sum_{i=1}^M \left \| f(\bm{x}_i) - \bm{c}_{y_i} \right \| ^2_2 
	\end{split}
	\end{equation}
\end{scriptsize}
	The $\bm{c}_{y_i}  \in \mathbb{R}^d $ denotes the $y_{i}$th class center of deep features. The formulation
	effectively characterizes the intra-class variations. A scalar $\lambda$ is used for balancing the two loss functions.
	
	The conventional softmax loss can be considered as a special case of this joint
	supervision, if $\lambda$ is set to 0. With proper $\lambda$, the discriminative power of deep features can
	be significantly enhanced \cite{wen2016discriminative}.

	\subsubsection{Angular Softmax loss}
	 In  \cite{liu2017sphereface}, the authors propose a  natural way to learn angular margin. The angular softmax (A-Softmax) loss is defined as 
	\begin{footnotesize}
	\begin{equation}
	\ell = \frac{1}{M}\sum_{i=1}^{M}-\log(\frac{e^{\left \| f(\bm{x}_i) \right \| \phi(\theta_{y_i}, i)}}{e^{\left \| f(\bm{x}_i) \right \| \phi(\theta_{y_i}, i)} + \sum_{j\neq y_i} e^{\left \| f(\bm{x}_i) \right \| cos(\theta_j, i)}})
	\end{equation}
	\end{footnotesize}
	where $\phi (\theta_{y_i}, i) = (-1)^k cos(m\theta_{y_i, i}) -2k$, $\theta(y_i, i) \in \left [ \frac{k\pi}{m}, \frac{(k+1)\pi}{m} \right ]$ and  $k \in \left [0, m - 1 \right ]$. $m \geq 1$ is an integer that controls the size of angular margin. When $m=1$, it becomes the modified softmax loss.  
	
A-Softmax loss has clear geometric interpretation. Supervised by A-Softmax loss, the learned features construct a discriminative angular distance metric that is equivalent to geodesic distance on a hypersphere manifold, which intrinsically matches the prior that speakers also lie on a manifold. A-Softmax loss has stronger requirements for a correct classification when $m \geq 2$, which generates an angular classification margin between learned features of different classes \cite{liu2017sphereface}.

	
	
	
	
	\section{Experiments}
	\label{sec:experiments}
	
	\subsection{Data description}

	\subsubsection{Voxceleb}
	Voxceleb is a large scale text-independent SR dataset collected ``in the wild", which contains over 100,000 utterances from 1251 celebrities. It can be used for both speaker identification and verification \cite{Nagrani17}. We pool the official split training and validation set together as our development dataset.

	For speaker verification task,  there are totally  1211 celebrities in the development dataset. The testing dataset contains 4715 utterances from the rest 40 celebrities. There are totally 37720 pairs of trials including 18860 pairs of  true trials. Two key performance metrics $C_{det}$ \cite{nist2012} and EER are used to evaluate the system performance for the verification task as shown in Table \ref{table:sv_table}. 
	
	For speaker identification task, there are totally  1251 celebrities  in the development dataset. The testing dataset contains 8251 utterances from these 1251 celebrities.  We report top-1 and top-5 accuracies as in Table \ref{table:sid_table}.

	\subsubsection{NIST LRE07}
The whole training corpus including  Callfriend datasets,  LRE 2003, LRE 2005, SRE 2008  datasets and development data for LRE07. The total training data is about 37000 utterances. The task of interest is the closed-set language detection. There are totally 14 target languages in testing corpus, which included 7530 utterances split among three nominal durations: 30, 10 and 3 seconds. Two key performance metrics Average Detection Cost  $C_{avg}$ \cite{nist2007} and Equal Error Rate (EER) are used to evaluate system performance as shown in Table \ref{table:lre07_table}. 

		\begin{table}[tb]
	\normalsize
	\centering
	\caption{Our end-to-end  baseline network configuration}
	\label{table:resnetconfig}
	\resizebox{0.98\columnwidth}{!}{
		\renewcommand\arraystretch{1.49}
		\begin{tabular}{|c|c|c|c|c|}
			\hline
			Layer               & Output size            & Downsample      &  Channels     &  Blocks      \\ \hline
			Conv1                     & $64$ $\times$ $L_{in}$                & False& 16       & -    \\ \hline
			Res1                & $64$ $\times$  $L_{in}$               & False& 16& 3    \\ \hline
			Res2                & 32 $\times$ $\frac{L_{in}}{2}$               & True& 32& 4    \\ \hline
			Res3                & 16 $\times$ $\frac{L_{in}}{4}$              & True& 64& 6   \\ \hline
			Res4                & 8 $\times$ $\frac{L_{in}}{8} $              & True& 128& 3    \\ \hline
			Avgpool          & 1 $\times$ $\frac{L_{in}}{8} $ &-&128&- \\ \hline
			Reshape         & 128$\times$  $L_{out}, L_{out}=\frac{L_{in}}{8}$ &-&-&- \\ \hline
	\end{tabular}}
\end{table}

\begin{table*} [tb] 
	\caption{   Results for verification on VoxCeleb (lower is better)}
	\centerline {
		\begin{tabular}{c c c c c  c c c c} 
			\hline
			\hline
			{ System ID}&  System Description&Encoding Procedure&Loss Function&  Similarity Metric&$C_{det}$&$EER(\%)$\\
			\hline	
			\hline
			1&i-vector + cosine   & Supervector&  GNLL&cosine&0.829&20.63\\
			2&i-vector + PLDA   & Supervector& GNLL +   GNLL&PLDA&0.639&7.95\\
			\hline	
			\hline
			3& TAP-Softmax  &TAP&softmax&cosine&0.553&5.48\\	
			4& TAP-Softmax  &TAP&softmax + GNLL&PLDA&0.545&5.21	\\	  
			\hline	
			5& TAP-CenterLoss  &TAP&center loss&cosine&0.522&4.75\\
			6& TAP-CenterLoss  &TAP&center loss+ GNLL&PLDA&0.5155&4.59\\ 
			\hline	
			7& TAP-ASoftmax   &TAP&A-Softmax&cosine&\textbf{0.439}&5.27\\				
			8& TAP-ASoftmax  &TAP&A-Softmax + GNLL&PLDA&0.577&\textbf{4.46}\\  
			\hline
			\hline
			9& SAP-Softmax  &SAP&softmax&cosine&0.522&5.51\\	
			10& SAP-Softmax  &SAP&softmax + GNLL&PLDA&0.545&5.08	\\	
			\hline	
			11& SAP-CenterLoss  &SAP&center loss&cosine&0.540&4.98\\
			12& SAP-CenterLoss  &SAP&center loss+ GNLL&PLDA&0.571&4.89\\ 
			\hline	
			13& SAP-ASoftmax   &SAP&A-Softmax&cosine&0.509&4.90\\				
			14& SAP-ASoftmax  &SAP&A-Softmax + GNLL&PLDA&0.622&4.40\\  
			\hline
			\hline
			15& LDE-Softmax  &LDE&softmax&cosine&0.516&5.21\\
			16& LDE-Softmax  &LDE&softmax + GNLL&PLDA&0.519&5.07\\
			\hline
			17& LDE-CenterLoss  &LDE&center loss&cosine&0.496&4.98\\
			18& LDE-CenterLoss  &LDE&center loss + GNLL&PLDA&0.632&4.87\\
			\hline	
			19& LDE-ASoftmax  &LDE&A-Softmax&cosine&\textbf{0.441}&\textbf{4.56}\\
			20& LDE-ASoftmax  &LDE&A-Softmax + GNLL&PLDA&0.576&4.48\\
			\hline
			\hline
	\end{tabular}}
	\label{table:sv_table}
\end{table*}

\begin{table} [!tb] 
	\caption{ Results for identification on VoxCeleb (higher is better)}
	\centerline{
		\resizebox{0.49\textwidth}{!}{
			\small
			\begin{tabular}{c c c c }
				\hline	
				{ System ID }&  System Description&Top-1$(\%)$&Top-5$(\%)$\\
				\hline
				1&i-vector + LogReg &65.8 &81.4\\
				\hline
				2& CNN-TAP & 88.5&94.9\\	
				3& \textbf{CNN-SAP} & 89.2&94.1\\	
				4& \textbf{CNN-LDE} & \textbf{89.9	}&\textbf{95.7}\\
				\hline
	\end{tabular}}}
	\label{table:sid_table}
\end{table}	

\begin{table} [tb] 
	\caption{  Performance on the 2007 NIST LRE closed-set task (lower is better)}
	\centerline{
		\resizebox{0.49\textwidth}{!}{
			\begin{tabular}{c c c c c}
				\hline	
				{ System}& \multirow{2}{*}{ System Description}&\multicolumn{3}{c}{$C_{avg}(\%)/EER(\%)$}\\
				\cline{3-5}
				ID&&3s&10s&30s\\
				\hline
				1&i-vector + LogReg  &20.46/17.71&8.29/7.00&3.02/2.27\\
				\hline
				2& CNN-TAP  &9.98/11.28&3.24/5.76&1.73/3.96\\
				3& \textbf{CNN-SAP}  &8.59/9.89&\textbf{2.49}/4.27&\textbf{1.09}/2.38\\	
				4& \textbf{CNN-LDE} & \textbf{8.25/7.75}&2.61/\textbf{2.31}&1.13/\textbf{0.96}\\
				\hline
	\end{tabular}}}
	\label{table:lre07_table}
\end{table}

	\subsection{i-vector system}
For general usage, we focus on the comparison on those systems that do not require  additional transcribed speech data and extra DNN acoustic model. 
	
	As for the  baseline  i-vector system, raw audio is converted to 7-1-3-7 based 56 dimensional SDC feature for LR task. For SR task, 20 dimensional  MFCC is augmented with their delta and double delta coefficients, making 60 dimensional MFCC feature vectors. A frame-level energy-based voice activity detection (VAD) selects features corresponding to speech frames. A 2048 components full covariance GMM UBM is trained, along with a 600 dimensional i-vector extractor.
	
	For closed-set speaker/language identification, a multi-class LogReg is adopted as the back-end classifier. For open-set verification, cosine similarity or PLDA with full rank is adopted. 
	
	\subsection{End-to-end system}
	Audio is converted to 64-dimensional Fbank with a frame-length of 25 ms, mean-normalized over a sliding window of up to 3 seconds. The same VAD processing as in i-vector baseline system is used here. 
	We  fix the front-end deep CNN module based on the well known ResNet-34 architecture \cite{He2016Deep}. The detail architecture is described in Table \ref{table:resnetconfig}. The total parameters of the front-end feature extractor is about 1.35 million. 
	
	In CNN-TAP system, a simple average pooling layer is built on top of the front-end CNN. In CNN-LDE system,  the TAP layer is replaced with a LDE layer. The number of dictionary components in CNN-LDE system is 64. 

The lose weight parameter $\lambda$ of center loss is set to 0.001 in our experiments. 
For A-Softmax loss, we use the angular margin $m = 4$.

	The model is trained with a mini-batch, whose  size varies from 96 to 256 considering different datasets and model parameters. The network is trained using typical stochastic gradient descent with momentum 0.9 and weight decay 1e-4.  The learning rate is set to 0.1, 0.01,
	0.001 and is switched when the training loss plateaus. The training is finished at 40 epochs for Voxceleb dataset and 90 epochs for LRE 07 dataset. Since we have no separated validation set, the converged model after the last optimization step is used for evaluation.  For each training step,  an integer $L$ within $\left[ 300  \textrm{,}  800 \right]$  interval is randomly generated, and each data in the mini-batch is cropped or extended to $L$ frames. 
	
	For open-set speaker verification, the 128-dimensional speaker embedding is extracted after the penultimate layer of neural network. Additional similarity metric like cosine similarity or PLDA is adopted to generate the final pairwise score.

	In the testing stage, all the testing utterances with different duration are tested on the same model. Since the duration is arbitrary, we feed the testing speech utterance to the trained neural network one by one.

	\subsection{Evaluation}
As expected, the end-to-end learning systems outperform the conventional i-vector approach significantly for both SR and LR tasks (see Table 2-4).

For encoding layer, as can be observed in Table 1-3, both SAP layer and LDE layer outperform the baseline TAP layer. Besides,  the LDE layer system also show superior performance compared with SAP layer. Considering loss functions in Table 1,  in most cases,  systems trained with discriminative loss function like center loss or A-Softmax loss achieve better results than softmax loss.  In terms of similarity metric, we can find that PLDA gets significant error reduction in conventional i-vector approach. However, when it turns into end-to-end system, especially for those system trained with discriminative loss funtion, PLDA achieves little gain  and sometimes makes the result worse. 

Finally, CNN-LDE based end-to-end systems achieve best result in speaker/language identification task. Compared with CNN-TAP baseline system, the CNN-LDE system achieve 25\%, 45\%, 63\% relative error reduction for corresponding NIST LRE 07 3s, 10s, 30s duration task.  For Voxeceleb speaker identification task, system trained with LDE layer get relative 12\% error reduction compared with CNN-TAP system.

In speaker verification task, the speaker embeddings extracted from neural network trained in LDE-ASoftmax system perform best. In the testing stage, a simple cosine similarity achieves the result of $C_{det}$ 0.441 and EER 4.56\%, which achieves relative 20\% error reduction compared with TAP-Softmax baseline system. 

	\section{Conclusions}
	\label{sec:conclusions}
	In this paper, a unified and interpretable end-to-end system is developed for both SR and LR. It accepts variable-length input and produces an utterance level result. We investigate how to enhance the system by exploring different kinds of  encoding layers  and loss function. Besides the basic TAP layer, we introduce a SAP layer and a LDE layer to get the utterance level representation. In terms of loss function for open-set speaker verification,  center loss and A-Softmax loss is introduced to get more discriminative speaker embedding. Experimental results show that the performance of end-to-end learning system could be significantly improved by designing suitable encoding layer and loss function.
\section{Acknowledgement}  
The authors would like to acknowledge Yandong Wen from Carnegie Mellon University. He gives insightful advice on the implementation of end-to-end discriminative loss. 

This research was funded in part by the National Natural Science Foundation of China (61401524,61773413), Natural Science Foundation of Guangzhou City (201707010363), Science and Technology Development Foundation of Guangdong Province (2017B090901045), National Key Research and Development Program (2016YFC0103905).

	\newpage

	\bibliographystyle{IEEEbib}
	\bibliography{camera_ready}

	%

\end{document}